\documentstyle[12pt,prd,aps,epsfig,preprint]{revtex}
\begin{document}
\draft
\date{\today}
\title{One Dimensional Schr\"{o}dinger Equation
 With Two Moving Boundaries}

\author{Ercan Y{\i}lmaz~\thanks{yercan@metu.edu.tr}}
\address{ {\it Physics Department, Middle East Technical University,
06531 Ankara, Turkey}}
\maketitle

\begin{abstract}

In this letter, we consider the Schr\"{o}dinger equation for a
well potential with varying width. We solve one dimensional
Schr\"{o}dinger equation subject to time-dependent boundary
conditions for a spinless particle inside infinite potential well,
both wall of which move opposite direction with different
velocities $\upsilon_{1}$ and $\upsilon_{2}$, respectively.
\end{abstract}

\thispagestyle{empty} ~~~~\\ 
\newpage
\setcounter{page}{1}
\section{Introduction}

A well-known feature of the Schr\"{o}dinger equation is its
property of Galilean covariance. In section II, we briefly review
Galilean covariance and give the transformation properties
\cite{R1}. Recently, Makowski et al.\cite{R2,R3,R4} have reported
exact solutions for the schr\"{o}dinger equation submitted to time
dependent boundary conditions. The case they solved describe a
particle bouncing between two infinitely walls, one wall of which
is fixed and the other is allowed to move according to a function
L(t). To our knowledge, most of attempts have been focused on
solution of the Schr\"{o}dinger equation for a particle between
two infinitely walls, but one of which is fixed \cite{R5,R6}.

There are two motivations for this study. The first one, we will
see the group of transformations admitted by the problem. In
quantum mechanics, one of the main task is to solve the
Schr\"{o}dinger equation. The second motivation is physical; the
solutions of the Schr\"{o}dinger equation for one-dimensional
system with two moving boundaries as seen in Fig.\ref{fig:fig1}
are not cited in literature and this subject is worthy to be
studied \cite{R7}. We solve the Schr\"{o}dinger equation for a
particle between two infinitely walls, but both of which move in
opposite directions. If the velocities of the moving walls are
constant, there exits a set of exact solutions for any values of
the velocities of the moving walls \cite{R8}.

\section{GALILEAN COVARIANCE IN THE CASE OF A SCALAR POTENTIAL}

We assume there exists some inertial frame S, relative to which
the Schr\"{o}dinger equation is valid for a spinless particle in
the presence of a scalar potential V. The Schr\"{o}dinger equation

\begin{eqnarray}\label{e1}
i\hbar\frac{\partial\psi(x,t)}{\partial
t}=\left[\frac{1}{2m}(-i\hbar\nabla)^{2}+V(x,t)\right]\psi(x,t).
\end{eqnarray}
is to be solved under Dirichlet boundary conditions \cite{R2}.
Relative to a new inertial frame $S'$ moving at velocity
$\upsilon_1$ with respect to S, the value of the wave function at
an arbitrary space-time location is related to that of $\psi$ at
the same location by a phase factor to ensure invariance of the
probability density at that position:
\begin{eqnarray}\label{e2}
\psi(x,t)=e^{-i\phi}\psi'(x',t').
\end{eqnarray}
In general, it is difficult to work out the exact solutions for a
moving boundary system due to the moving-boundary conditions
\begin{eqnarray}\label{e3}
\psi(-\upsilon_{1} t, t)=0~~~~ and ~~~~ \psi(a+\upsilon_{2} t,
t)=0
\end{eqnarray}
 It is well-known that Eq. (1) is covariant under the following
transformations:
\begin{eqnarray}\label{e4}
x'=x-\upsilon_{1} t,~~~~~~ t=t',
\end{eqnarray}

\begin{eqnarray}\label{e5}
\frac{\partial}{\partial x}=\frac{\partial}{\partial
x'},~~~~~~\frac{\partial}{\partial t}= \frac{\partial}{\partial
t'}-\upsilon_{1}\frac{\partial}{\partial x'}
\end{eqnarray}

\begin{eqnarray}\label{e6}
V(x,t)=V'(x',t')
\end{eqnarray}
We consider the case the potential V=0. As we know, under the
Galilean transformations the time derivative of $\psi$ and
$\frac{\partial^{2}}{\partial x^{2}}$ will be changed as follows:
\begin{eqnarray}\label{e7}
\dot{\psi}(x,t)=-i\dot{\phi}e^{-i\phi}\psi'(x',t')+e^{-i\phi}\dot{\psi'}(x',t')
\end{eqnarray}
Where $\dot{\phi}=\frac{\partial\phi}{\partial t'}$~~~ and
~~~$\dot{\psi}=\frac{\partial\psi}{\partial t'}$ and
\begin{eqnarray}\label{e8}
\frac{\partial^{2}}{\partial
x^{2}}\psi(x,t)=e^{-i\phi}[-i\frac{\partial^{2}\phi}{\partial
x'^{2}}-(\frac{\partial\phi}{\partial x'})^{2}]
\psi'(x',t')-2ie^{-i\phi}\frac{\partial\phi}{\partial
x'}\frac{\partial\psi'(x',t')}{\partial
x'}+e^{-i\phi}\frac{\partial^{2}\psi'(x',t')}{\partial x'^{2}}
\end{eqnarray}
We get the Schr\"{o}dinger equation under the Galilean
transformation;
\begin{eqnarray}\label{e9}
-\frac{\hbar^{2}}{2m}\frac{\partial^{2}\psi'(x',t')}{\partial
x'^{2}}+(\frac{i\hbar^{2}}{m}\frac{\partial\phi}{\partial
x'}+i\hbar\upsilon_{1})\frac{\partial\psi'(x',t')}{\partial x'}
 \nonumber \\+
[\frac{\hbar^{2}}{2m} (\frac{\partial\phi}{\partial
x'})^{2}+\frac{i\hbar^{2}}{2m}\frac{\partial^{2}\phi}{\partial
x'^{2}}-\hbar\frac{\partial\phi}{\partial
t'}+\hbar\upsilon_{1}\frac{\partial\phi}{\partial x'}]
\psi'(x',t') =i\hbar\frac{\partial\psi'(x',t')}{\partial t'}
\end{eqnarray}
The problem posed in this way is unsolvable, so we define a new
coordinate as given in section III.

\section{FUNDAMENTAL TRANSFORMATION}

The main purpose is to change the unsolvable moving boundary
problem into a solvable one side fixed-boundary problem as seen in
Fig.\ref{fig:fig2} by using the following transformation.

Let us define a new rescaled space coordinate and from now on we
use t in stead of $t'$.
\begin{eqnarray}\label{e10}
\bar{x}=\frac{x'}{L(t')}
\end{eqnarray}
where $L(t')=\left[1+\frac{(\upsilon_1+\upsilon_2)t'}{a}\right]$.
Than the derivative can be written as follows:
\begin{eqnarray}\label{e11}
\frac{\partial}{\partial x'}=\frac{\partial\bar{x}}{\partial
x'}\frac{\partial}{\partial\bar{x}}=\frac{1}{L(t')}\frac{\partial}{\partial\bar{x}}
\end{eqnarray}
and
\begin{eqnarray}\label{e12}
\frac{\partial^{2}}{\partial
x'^{2}}=\frac{1}{L^{2}(t')}\frac{\partial^{2}}{\partial\bar{x}^{2}}
\end{eqnarray}
We have got time derivative
\begin{eqnarray}\label{e13}
\frac{\partial}{\partial
t'}\longrightarrow\frac{\partial}{\partial
t'}-\frac{\bar{x}\dot{L}(t')}{L(t')}\frac{\partial}{\partial\bar{x}}
\end{eqnarray}

The new form of Schr\"{o}dinger equation by using these
transformations and substituting $x'=\bar{x}L(t')$ is
\begin{eqnarray}\label{e14}
-\frac{\hbar^{2}}{2m}\frac{1}{L^{2}}\frac{\partial^{2}\bar{\psi}}{\partial
\bar{x}^{2}}+
[\frac{i\hbar^{2}}{mL^{2}}\frac{\partial\bar{\phi}}{\partial
\bar{x}}+\frac{i\hbar\upsilon_{1}}{L}
+\frac{i\hbar\bar{x}\dot{L}}{L}]\frac{\partial\bar{\psi}}{\partial
\bar{x}}
 \nonumber\\+
 [\frac{\hbar^{2}}{2mL^{2}}(\frac{\partial\bar{\phi}}{\partial
 \bar{x}})^{2}+
 \frac{i\hbar^{2}}{2mL^{2}}\frac{\partial^{2}\bar\phi}{\partial
 \bar{x}^{2}}-
 \hbar\frac{\partial\bar\phi}{\partial t'}+
 \hbar\frac{\bar{x}\dot{L}}{L}\frac{\partial\bar\phi}{\partial \bar{x}}+
 \frac{\hbar\upsilon_{1}}{L}\frac{\partial\bar\phi}{\partial \bar{x}}]\bar\psi=
 i\hbar\frac{\partial\bar\psi}{\partial t'}
\end{eqnarray}
the second term and third term of Eq. (14) equal zero, separately.
So from the second term it is easy to find $\bar\phi$ as follows;
\begin{eqnarray}\label{e15}
[\frac{i\hbar^{2}}{mL^{2}}\frac{\partial\bar\phi}{\partial
\bar{x}}+\frac{i\hbar\upsilon_{1}}{L}
+\frac{i\hbar\bar{x}\dot{L}}{L}]=0
\end{eqnarray}

\begin{eqnarray}\label{e16}
\bar\phi=-\frac{m}{\hbar}[\frac{\bar{x}^{2}}{2}\dot{L}+\upsilon_{1}\bar{x}]L+f(t')
\end{eqnarray}
One can easily calculate
${\bar\phi}''=\frac{\partial^{2}\bar\phi}{\partial \bar{x}^{2}}$
and $\dot{\bar\phi}=\frac{\partial\bar\phi}{\partial t'}$ as
follows;
\begin{eqnarray}\label{e17}
\bar{\phi}''=-\frac{m}{\hbar}\dot{L}L
\end{eqnarray}
and
\begin{eqnarray}\label{e18}
\dot{\bar\phi}=-\frac{m}{\hbar}[\frac{\bar{x}^{2}}{2}\dot{L}+\upsilon_{1}\bar{x}]\dot{L}
-\frac{m}{\hbar}L\ddot{L}\frac{\bar{x}^{2}}{2}+\dot{f}(t')
\end{eqnarray}
The second term of Eq. (18) will be zero because of $\ddot{L}=0$.
The third term of Eq. (14) can be rewritten as follows;

\begin{eqnarray}\label{e19}
\frac{1}{L^{2}}\{{\frac{i\hbar^{2}}{2m}\frac{\partial^{2}\bar\phi}{\partial
\bar{x}^{2}}+\frac{\hbar^{2}}{2m}(\frac{\partial\bar\phi}{\partial
\bar{x}})^{2}
+\hbar(\upsilon_{1}+\bar{x}\dot{L})L\frac{\partial\bar\phi}{\partial
\bar{x}}-\hbar L^{2}\frac{\partial\bar\phi}{\partial
t'}}\}\bar\psi=0
\end{eqnarray}
When we substitute Eq. (16,17,18) into Eq. (19), we get the
following equation;
\begin{eqnarray}\label{e20}
\frac{1}{2}mL^{2}\upsilon_{1}^{2}-\frac{i\hbar}{2}L\dot{L}- \hbar
L^{2}\dot{f}(t')=0
\end{eqnarray}
after manipulation we get,
\begin{eqnarray}\label{e21}
\dot{f}(t')=\frac{m}{2\hbar}\upsilon_{1}^{2}-\frac{i\dot{L}}{2L}
\end{eqnarray}
and it is easy to calculate $f(t')$ by using
$L=1+\frac{\upsilon_{1}+\upsilon_{2}}{a}t'$ and
\begin{eqnarray}\label{e22}
f(t')=\frac{m}{2\hbar}\upsilon_{1}^{2}t'-\frac{i\dot{L}a}
{2}\int\frac{dt'}{a+(\upsilon_{1}+\upsilon_{2})t'}
\end{eqnarray}
The remain terms in Eq. (14) which are different zero are
following;
\begin{eqnarray}\label{e23}
-\frac{\hbar^{2}}{2m}\frac{1}{L^{2}}\frac{\partial^{2}\bar\psi}{\partial
\bar{x}^{2}} =i\hbar\frac{\partial\bar\psi}{\partial t'}
\end{eqnarray}
Eq. (23) must be equal to a constant in order to be solved,
\begin{eqnarray}\label{e24}
-\frac{\hbar^{2}}{2m}\frac{\partial^{2}\bar\psi}{\partial
\bar{x}^{2}} =i\hbar L^{2}\frac{\partial\bar\psi}{\partial
t}=E\bar\psi
\end{eqnarray}
where E is energy. By using separation of variables we get
\begin{eqnarray}\label{e25}
\bar\psi(\bar{x},t')=X(\bar{x})T(t')
\end{eqnarray}
space part of equation Eq. (24) can be rewritten as follows,
\begin{eqnarray}\label{e26}
-\frac{d^{2} X(\bar{x})}{\partial\bar{x}^{2}}=k^{2}X(\bar{x})
\end{eqnarray}
where $k^{2}=\frac{2mE}{\hbar^{2}}$ and the solution
\begin{eqnarray}\label{e27}
X(\bar{x})=A\sin k\bar{x}+B\cos k\bar{x}
\end{eqnarray}
and the time-dependent part of Eq. (24) can be rewritten as
follows,
\begin{eqnarray}\label{e28}
i\hbar L^{2}\dot{T}(t')=ET(t')
\end{eqnarray}
and solution of time part is:
\begin{eqnarray}\label{e29}
T(t')=\exp(-\frac{iE}{\hbar}\int\frac{dt'}{L^{2}(t')})
\end{eqnarray}
one can easily found the solution:
\begin{eqnarray}\label{e30}
\bar\psi(\bar{x},t')=\exp\left[-\frac{iE}{\hbar}\int\frac{dt'}{L^{2}(t')}\right][A\sin
k\bar{x}+B\cos k\bar{x}]
\end{eqnarray}
The function (30) is a correct solution of Eq. (14) for both
casses if it vanishes at $\bar{x}=0$ and $\bar{x}=a$. For
$\bar{x}=0$ the function $\bar{\psi}(\bar{x},t')=0$ is obviously.
To fulfil the second boundary condition the constant k have to be
chosen in such a way that $\bar{\psi}(\bar{x},t')=0$ when
$\bar{x}=a$.

Under Eq. (4 and 10) transformation, boundary conditions in Eq.
(3) transform as follows

\begin{eqnarray}\label{e31}
\bar\psi(0, t)=0~~~~ and ~~~~\bar\psi(a, t)=0
\end{eqnarray}

k can be find by using boundary conditions,
\begin{eqnarray}\label{e32}
at~~~~ \bar{x}=0,~~~
\bar\psi(\bar{x},t')=\psi'(\frac{x'}{L(t')},t')=0\Longrightarrow
B=0
\end{eqnarray}
and
\begin{eqnarray}\label{e33}
at~~~~ \bar{x}=a,~~~ \bar\psi(\bar{x},t')=0\Longrightarrow
k=\frac{n\pi}{a}
\end{eqnarray}
finally, the exact solution of our problem is
\begin{eqnarray}\label{e34}
\psi(x,t)=A\exp\left[-\frac{iE}{\hbar}\int\frac{dt}{L^{2}}\right]\sin\left[\frac{n\pi}{a}
\frac{(x-\upsilon_1 t)}{1+\frac{(\upsilon_1
+\upsilon_2)t}{a}}\right]e^{-i\phi}
\end{eqnarray}
where A is a normalization constant and, $\phi$ and f(t) are
follows, respectively.
\begin{eqnarray}\label{e35}
\phi=-\frac{m}{\hbar}\left[\frac{1}{2}\frac{(x-\upsilon_{1}t)^{2}}{L}\dot{L}
+\upsilon_{1}(x-\upsilon_{1}t)\right]+f(t)
\end{eqnarray}
\begin{eqnarray}\label{e36}
f(t)=\frac{m}{2\hbar}\upsilon_{1}^{2}t-\frac{i\dot{L}a}
{2}\int\frac{dt}{a+(\upsilon_{1}+\upsilon_{2})t}
\end{eqnarray}

\section{CONCLUSION}
As a conclusion of this study, we notice that the general problem
becomes much simpler by using fundamental transformation. However,
the moving-boundary condition renders the equation unsolvable by
the usual means. The reader can easily be convinced of the
difficulty of this problem by trying a method of separation of
variables on Eq. (9) without using the fundamental transformation.
It can be easily seen that this fundamental transformation leads
to an ordinary differential equation, the solution of which can be
achieved by using separation of variables.

\section{Acknowledgements}
The author is grateful to Cem Yuce for helpful discussions.

\begin{figure}[h]
\vspace{1cm} \includegraphics{pot.ps} \vspace{6cm} \caption{Potential
well with two moving boundaries. \label{fig:fig1} }
\end{figure}

\begin{figure}[h]
\vspace{1cm} \includegraphics{pot2.ps} \vspace{6cm} \caption{Potential
well with one fixed-wall and one moving wall. \label{fig:fig2} }
\end{figure}


\begin{references}
\bibitem{R1} H. R. Brown and P. R. Holland, Am. J. Phys. {\bf 67(3)} (1999) 204.
\bibitem{R2} A. J. Makowski and S. T. Dembinski, Phys. Lett. {\bf A 154} (1991) 217.
\bibitem{R3} A. J. Makowski and P. Peplowski, Phys. Lett. {\bf A 163} (1992) 142.
\bibitem{R4} A. J. Makowski, J. Phys. {\bf A 25} (1992) 3419.
\bibitem{R5} D. A. Morales, Z. Parra and R. Almeida, Phys. Lett. {\bf A 185} (1994) 273.
\bibitem{R6} A. Munter, J. R. Burgan, M. Feix and E. Fijalkow, J. Math. Phys. {\bf  22(6)} (1981) 1219.
\bibitem{R7} L. Li and B. Z. Li, Phys. Lett. {\bf A 291} (2001) 190.
\bibitem{R8} S. W. Doescher and M. H. Rice, Am. J. Phys. {\bf 37(12)} (1969) 1246.
\end{references}
\end{document}